# Enhancing plasticity in high-entropy refractory ceramics via tailoring valence electron concentration

Davide G. Sangiovanni[1,†], William Mellor[2,3,†], Tyler Harrington[3], Kevin Kaufmann[2], and Kenneth Vecchio[2,3,*]

[1]Department of Physics, Chemistry, and Biology (IFM) Linköping University, SE-581 83, Linköping, Sweden

[2]Department of NanoEngineering, UC San Diego, La Jolla, CA 92093, USA

[3]Materials Science and Engineering Program, UC San Diego, La Jolla, CA 92093, USA

† *These authors contributed equally to this work*

\* *Corresponding author:* kvecchio@eng.ucsd.edu

## ABSTRACT

Bottom-up design of high-entropy ceramics is a promising approach for realizing materials with unique combination of high hardness and fracture-resistance at elevated temperature. This work offers a simple yet fundamental design criterion – valence electron concentration (VEC) ⪆ 9.5 e$^-$/f.u. to populate bonding metallic states at the Fermi level – for selecting elemental compositions that may form rocksalt-structure (B1) high-entropy ceramics with enhanced plasticity (reduced brittleness). Single-phase B1 (HfTaTiWZr)C and (MoNbTaVW)C, chosen as representative systems due to their specific VEC values, are here synthesized and tested. Nanoindentation arrays at various loads and depths statistically show that (HfTaTiWZr)C (VEC=8.6 e$^-$/f.u.) is hard but brittle, whilst (MoNbTaVW)C (VEC=9.4 e$^-$/f.u.) is hard *and* considerably more resistant to fracture than (HfTaTiWZr)C. *Ab initio* molecular dynamics simulations and electronic-structure analysis reveal that the improved fracture-resistance of (MoNbTaVW)C subject to tensile and shear deformation may originate from the intrinsic material's ability to undergo local lattice transformations beyond tensile yield points, as well as from relatively facile activation of lattice slip. Additional simulations, carried out to follow the evolution in mechanical properties as a function of temperature, suggest that (MoNbTaVW)C may retain good resistance to fracture up to ≈900-1200K, whereas (HfTaTiWZr)C is predicted to remain brittle at all investigated temperatures.

Keywords: High entropy ceramics, *ab initio* molecular dynamics, stress-induced transformation, fracture toughness, nanoindentation, valence electron concentration

# 1. Introduction

The rapidly expanding field of high entropy alloys has led to the recent development of novel material systems with remarkable or unprecedented properties [1–10]. Following the trend of high entropy alloys, the theory of configurational entropy stabilization and exploration of compositionally complex solid solutions has carried over to ceramics, lending to the ability to now properly identify plausible compositions previously thought unobtainable. Rost *et al.* [5,9,10] reported the first entropy-stabilized oxide, a rock-salt (B1) structured (MgCoNiCuZn)O, which demonstrated formation of a single phase above a critical temperature. Since this initial work, many reports on high entropy ceramics including oxides [11,12], borides [13–16], carbides [6,8,17–23], nitrides [24–26], carbonitrides [18,26], and silicides [27,28] have been published. Of these material systems, high entropy carbides (HEC) are likely candidates for ultra-high temperature applications, such as ultra-sonic flight leading edges and high temperature reactor linings, due to their many attractive properties, including high melting temperatures and high-temperature mechanical stability [6,8,19,29]. HEC have demonstrated increased hardness [6,8] over their expected rule of mixtures values, as well as improved creep resistance [22], good oxidation resistance [30], low thermal conductivity [31,32], and high resistance to irradiation damage [33]. The high entropy carbide (HfNbTaTiZr)C has been shown to exhibit a high flexural strength of 421MPa at room temperature, which is maintained to 1800°C and decreases to 318MPa at 2000°C [34]. The high hardness has been attributed to the high degree of randomness and lattice distortions, making it unique to high entropy materials [21,34]. Nevertheless, high hardness (correlated to high mechanical strength in solids [35]) is often insufficient to prevent brittle fracture, a problem which severely limits the potential use of ceramics. In fact, the combination of high strength and excellent fracture resistance – which equates to superior toughness – is generally unattainable for a monolithic solid ceramic phase [36].

The vast composition space of high entropy ceramics offers the possibility to optimize a desired set of mechanical properties in a single-phase bulk material. However, predictions of strength, ductility, and toughness in high-entropy ceramics have been typically based upon phenomenological criteria, such as Pugh's shear-to-bulk moduli ratios and Cauchy's pressures derived from first-principles calculations [37,38]. Fundamental understanding of the relationship that links the composition of a high-entropy ceramic phase to its inherent toughness is currently missing. Nonetheless, recent theoretical and experimental investigations conducted on pseudobinary rock-salt structured (B1) refractory nitride ceramics showed that the elastic properties, plasticity mechanisms and, therefore, toughness of the material can be controlled through manipulation of the electronic structure [39-42]. More specifically, the toughness of pseudobinary refractory nitrides is enhanced with increasing valence electron concentration (VEC) [41,43,44] up to $\approx 10.5$ e$^-$/f.u.; a threshold beyond which the B1 structure may become unstable [41,45]. A VEC $\approx 10.5$ e$^-$/f.u. corresponds to full occupation of bonding metallic *d-d* states at the Fermi level, which renders the ceramic more compliant to shearing (improved ductility), while retaining high mechanical strength and hardness (schematic explanation in figure 3 of Ref. [42]). The rationale of improving the toughness of nitride ceramics via VEC-tuning has been validated via synthesis and mechanical testing of single-crystal B1 (V,Mo)N deposited as thin films on a substrate [46]. Other preliminary *ab initio* investigations suggested that an increased VEC may also be beneficial for the toughness of B1 pseudobinary transition-metal carbides [47]. However, an experimental verification of such predictions is lacking.

In comparison to pseudobinary systems, the problem of assessing toughness in high entropy carbides via experiments and calculations is considerably more complex due to the several degrees of freedom of multicomponent systems. A previous investigation by the present authors demonstrated that the hardness (H) and modulus (E) of high entropy carbides decrease and increase, respectively, with increasing VEC [6]. Phenomenological criteria based on H/E and $H^3/E^2$ values, proposed for predicting the resistance to fracture of hard ceramics [48,49], would (considering the experimental results of Ref. [6]) suggest that the overall toughness of HEC decreases with increasing VEC. Conversely, our recent *ab initio* molecular dynamics calculations and sound-wave speed measurements of shear and elastic moduli in the HEC (HfTaTiWZr)C and (MoNbTaVW)C, indicate that (MoNbTaVW)C (VEC=9.4 e$^-$/f.u.) is more ductile than (HfTaTiWZr)C (VEC=8.6 e$^-$/f.u.) [50]. This latter prediction, combined with the fact that the two HEC have comparable hardness (27-33 GPa) [50], would suggest that (MoNbTaVW)C may exhibit enhanced plasticity compared to (HfTaTiWZr)C, which is consistent with expectations based on the materials' VEC.

Here, we employ nanoindentation testing at constant loads and constant penetration depths, as well as *ab initio* molecular dynamics (AIMD) simulations at room temperature, to provide insights on the mechanisms behind the dependence of mechanical properties of high-entropy carbides on their composition and VEC, thus offering general search criteria for ceramic systems that combine high hardness and toughness. AIMD simulations, used to calculate the full stress/strain curves of (HfTaTiWZr)C and (MoNbTaVW)C subject to tensile and shear deformation, reveal atomistic pathways responsible for ductile fracture vs. brittle cleavage, as well as differences in shear strengths required to induce lattice slip on different crystallographic planes. The results of the simulations, also supported by electronic-structure analyses of highly-deformed supercells, allow for interpretation of the relative differences in fracture resistance of (HfTaTiWZr)C and (MoNbTaVW)C specimens observed during nanoindentation experiments. Moreover, given that refractory ceramics are of paramount importance in applications at elevated temperature, we carry out additional AIMD simulations of mechanical testing to investigate the evolution in mechanical properties of (HfTaTiWZr)C and (MoNbTaVW)C at 600, 900, and 1200K.

## 2. Methods
### 2.1 Sample Preparation

Two specific 5-cation, high entropy carbides were prepared for this study: (HfTaTiWZr)C and (MoNbTaVW)C. Individual carbide compounds (HfC, TaC, TiC, VC, W$_2$C, ZrC, and Mo$_2$C + graphite) were utilized as starting powders (Alfa Aesar, >99.5% purity, -325 mesh). Each carbide precursor was weighed out in equiatomic amounts, with the Mo$_2$C and W$_2$C amounts adjusted with additions of graphite to achieve the monocarbide compositions of MoC and WC. Subsequently, the powders were inserted into a tungsten carbide lined stainless steel jar containing 10mm tungsten carbide milling media, and high energy ball milled (HEBM) in a SPEX 8000D shaker pot high energy ball mill (SpexCerPrep, NJ, USA) for two hours in 12-gram batches. The entire two-hour milling process was divided into 4 cycles of 30 minutes with 15 minutes of rest after each individual cycle to allow for thermal cooling and to reduce oxide formation. The entire HEBM process was carried out in an inert argon atmosphere to further minimize possible oxidation of the powders. Following the HEBM process, the sample powders were encapsulated within 20mm graphite dies lined with graphite foil on all sides and readied for spark plasma sintering (SPS). The samples were spark plasma sintered using a Thermal Technologies 3000 series SPS (Thermal Technologies, CA, USA). Samples were sintered at 2473K with a heating rate of 100 K/min. A uniaxial pressure of 80 MPa was applied at the maximum temperature. The sample was left to slow cool in a vacuum environment until room temperature is reached. The resulting sample was a 20mm diameter specimen approximately 4 mm in thickness.

**2.2 X-ray diffraction and electron microscopy**

Crystal structure analysis was implemented using a Rigaku Miniflex X-ray Diffractometer with a 1D detector. Diffraction scans were carried out across a 2θ of 20°-120° using a step size of 0.02° and a scan rate of 5 degrees per minute. For all measurements, Copper $K_\alpha$ radiation (wavelength λ = 1.54059 Å) was used. XRD phase analysis is carried out utilizing Match!™ phase identification software.

Microstructural and elemental analysis was performed using a Thermo Scientific Apreo field emission scanning electron microscope (SEM) equipped with an Oxford X-Max$^N$ EDS detector and an Oxford Symmetry electron backscatter diffraction (EBSD) detector. A combination of both secondary and backscattered electron images was utilized to analyze nanoindentations at an accelerating voltage of 20 kV.

**2.3 Mechanical Property Testing**

Samples underwent nanoindentation on a KLA-Tencor G200 Nanoindenter (KLA-tencor, CA, USA). Prior to any indentation measurement, the sample densities were measured via Archimedes principle. The measured density to theoretical density ratio was determined using the lattice parameter determined from their respective x-ray diffraction patterns. Both samples were determined to be 99% dense ensuring that the observed mechanical behaviors were not influenced by microstructural and porosity effects. Hardness measurements were performed under ISO standard 1477 across the range of 50-500 mN in increments of 50mN; this resulted in 10 test loads. These test loads were divided into a minimum of 6 arrays totaling 30 indents at each load. The arrays were comprised of 10 columns of each test load, with 5 tests in each column, and a minimum batch of 30 indents were performed. Indents were spaced 50 µm apart to avoid residual stress effects from previous indents. The binary 95% confidence intervals for the fraction of indents exhibiting fracture was calculated using the Adjusted Wald method [51]. The Adjusted Wald method offers ease of computation, is practical for small sample sizes (n as low as 5), and results in confidence intervals that are not overly narrow or wide [52]. The purpose of performing the arrays of nanoindentations was to document the number of nanoindentations that result in observable cracks as both a function of testing load and indentation depth (equivalent strain). Large numbers of indentations at each test condition were performed to provide statistically significant results that exceed any standard deviation or random variations in observed cracking.

**2.4 *Ab Initio* Molecular Dynamics Modeling**

Born-Oppenheimer AIMD is performed using the VASP [53] software. The simulations employ the projector augmented-wave method [54] and the generalized gradient approximation of Perdew-Burke-Ernzerhof to model electronic exchange and correlation energies [55]. Γ-point sampling of the reciprocal space and cutoff energies of 300 eV are used in all simulations. The classical equations of motion are solved at timesteps of 1 femtosecond (fs), with an energy tolerance of $10^{-5}$ eV/supercell. Three B1 HEC supercell models of different crystallographic orientations – namely, with the [001], [110], and [111] axis parallel to the z Cartesian direction – are used to simulate tensile deformation along [001], [110], and [111], as well as shearing on $\{001\}\langle 1\bar{1}0\rangle$, $\{110\}\langle 1\bar{1}0\rangle$, $\{111\}\langle 1\bar{1}0\rangle$, and $\{111\}\langle 11\bar{2}\rangle$ slip systems. All supercell models are comprised of 576 atoms (288 metals and 288 carbon) and are equivalent to those illustrated in figure 1a-c of Ref. [56]. For convenience, HEC supercells with vertical [hkl] orientation are denoted below as HEC (hkl). Lattice disorder in the HEC is mimicked by randomly distributing the five metal elements on the cation sublattices.

The supercell equilibrium structural parameters are determined as a function of temperature T = 300, 600, 900, and 1200K via NPT [57] sampling (≈5 ps) of the configurational space, implemented with Langevin's thermostat and the barostat of Parrinello and Rahman [58]. NPT dynamics allows verifying that the defect-free models maintain B1-structure symmetry during equilibration at a desired temperature. Thus, NVT canonical sampling [59] based on Nose-Hoover thermal bath is employed to equilibrate the unstrained structures for five additional picoseconds (ps). During NVT dynamics, it is ensured that time-averaged $|\sigma_{xx}|$, $|\sigma_{yy}|$, and $|\sigma_{zz}|$ stress components in equilibrium (unstrained) B1 supercells are all ≲ 0.2 GPa. We note that AIMD simulations at 300K yield lattice parameters of 4.477 Å for (HfTaTiWZr)C and 4.370 Å for (MoNbTaVW)C [50], in excellent agreement with our experimental values [6].

The AIMD approach used to determine the stress/strain relationships during uniaxial and shear deformation has been detailed previously [56,60,61]. Briefly, the tensile or shear strain δ is sequentially incremented (2%), equilibrating the structures at each deformation state for 3 ps at the targeted temperature. Note that canonical AIMD sampling implies that the volume of the simulation box remains constant during each strain. Elongation and shear deformation of the crystal are increased up to the occurrence of fracture, or slip, of the lattice.

AIMD tensile testing orthogonal to (001), (110), and (111) crystal surfaces – which have the lowest energy of formation in B1 structured ceramics [62] – allows determining the ideal[1] tensile strength $\sigma_T$, modulus of resilience $U_R$, and toughness $U_T$ for different strain directions. $\sigma_T$ corresponds to the maximum vertical ($\sigma_{zz}$) stress withstood during elongation, while $U_T$ is evaluated from the area that underlies the full stress/strain curve. $U_R$ is defined as the energy density absorbed during elastic deformation. Note that $U_R \approx U_T$ for materials that cleave at their yield point. The modeling of shear deformation of dislocation-free supercells allows evaluating the intrinsic resistance of the crystal to slip (or formation of stacking faults) on different crystallographic planes. For B1 structured ceramics, typical slip planes are the {110} and {111}, whereas the Burgers vector is [1$\bar{1}$0]; we also investigate lattice slip on {001} planes. The ideal shear strength $\gamma_S$, determined in our simulations for each slip system, corresponds to the maximum shear stress withstood by a dislocation-free lattice before slip. Snapshots and videos [see **Supplemental Material (SM)** video files] of AIMD simulations are generated with the VMD software [63].

---

[1] The term *ideal* indicates properties of defect-free bulk structures.

Electron-transfer maps of atomic configurations extracted from AIMD trajectories are obtained by subtracting the electron densities of non-interacting atomic orbitals from density-functional theory (DFT) self-consistent charge density.

## 3. Results
### 3.1 Phase Determination

Two sample compositions, corresponding to (HfTaTiWZr)C and (MoNbTaVW)C, were processed utilizing the ball milling and spark plasma sintering techniques described earlier. **Fig. 1** depicts both the XRD structural characterization and EDS analysis. For both sample compositions, it is observed that the B1 rock-salt structure was the only phase formed, where the additional reflections at higher 2θ correspond to Kα$_2$ splitting. Furthermore, the EDS analysis of both samples reveal that they are fully homogeneous throughout.

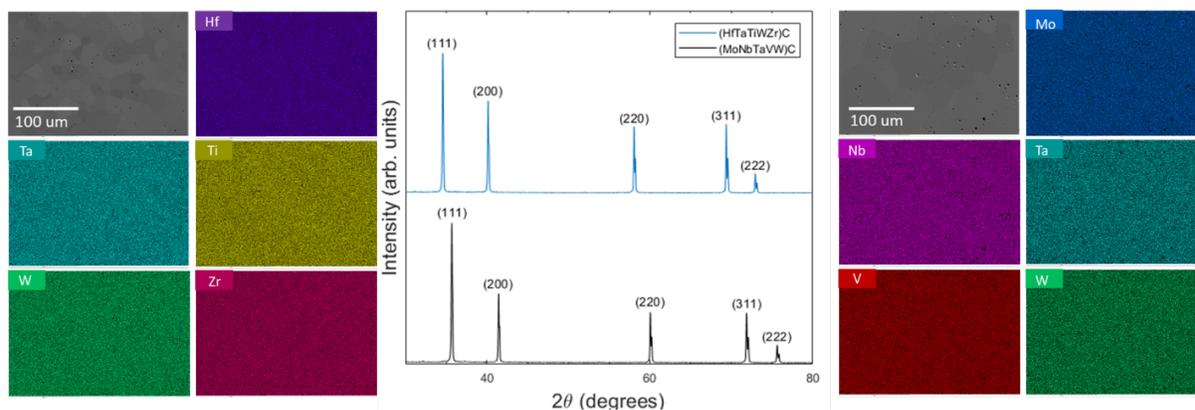

**Figure 1.** EDS maps of sample compositions (HfTaTiWZr)C (**left**) and (MoNbTaVW)C (**right**) with the XRD analysis of both (**middle**). For both compositions, the B1 rock-salt structure was observed with full homogeneity throughout each sample.

### 3.2 Nanoindentation Test Results

Since determination of fracture toughness of brittle materials by indentation crack measurements is highly variable at best, the relative toughness here is demonstrated in an alternate comparison. To determine the relative resistance of the individual carbides to fracture under an indentation load, the indentation load and depth at which fracture is detected around indents is recorded. **Fig. 2a** shows an example of an array of indents made with increasing load into a high entropy carbide sample. **Fig. 2b** shows the fraction of indents that exhibited detectable fractures for each indentation load in both (HfTaTiWZr)C and (MoNbTaVW)C. At an indentation load of 250 mN, 100 percent of indents in (HfTaTiWZr)C exhibit fracture, while it is only 33 percent for (MoNbTaVW)C. The fraction of indents exhibiting fracture for (HfTaTiWZr)C is 100 percent at any load in excess of 250 mN, while the (MoNbTaVW)C sample never reaches 100 percent of fractured indents up to 500 mN (the maximum tested in this study). **Fig. 2c** shows an example of a fractured indent in (HfTaTiWZr)C at a load of 450 mN and an unfractured indent in (MoNbTaVW)C at the equivalent load. Due to the difference in hardness for the two compositions, the indentation depth is not equivalent at the same load. To compare the fracture toughness considering relevant equivalent strains, **Fig. 2d** demonstrates the percent of indents that exhibit fracture at each measured indentation depth for both compositions. At an indentation depth of 600 nm, (HfTaTiWZr)C exhibits 100 percent fractured indents, while (MoNbTaVW)C shows only 30 percent. The 95% confidence intervals in **Fig. 2b,2d** were calculated using the Adjusted Wald method [51,52].

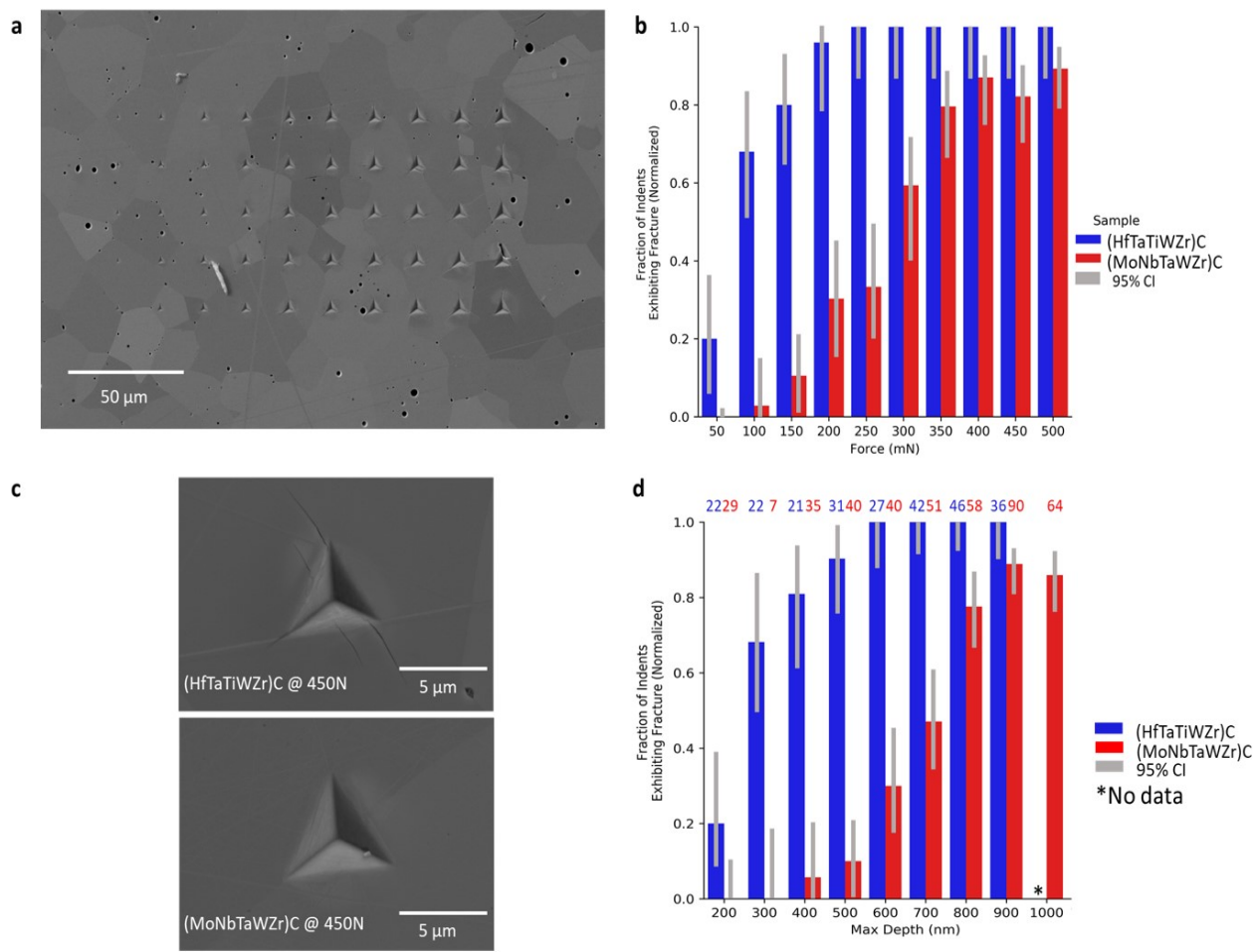

**Figure 2.** (a) Electron micrograph depicting a sample array. Indents here are spaced closer than reported tests to properly depict a test array. The spacing between indents in the actual tests was 50μm. (b) Test results from all arrays in both samples as a function of applied force. (c) Typical indents of each composition at a load of 450 mN, where fractures are clearly exhibited around the indent in (HfTaTiWZr)C and are absent in (MoNbTaVW)C. (d) The fraction of indents exhibiting fracture for each indentation depth.

### 3.3. *Ab Initio* Molecular Dynamics Results

Given that the nanoindentations are on the scale of 5μm in size, the fact that the average grain sizes in our HEC specimens are ~32μm for (HfTaTiWZr)C and ~43μm for (MoNbTaVW)C lends confidence that the relative resistance to fracture assessed via nanoindentation for the two carbide systems is negligibly affected by grain boundary properties. Accordingly, AIMD tensile-testing simulations carried out for defect-free single crystals provide an understanding for potential new plasticity routes for B1 (MoNbTaVW)C in relation to B1 (HfTaTiWZr)C. The stress/strain curves of (HfTaTiWZr)C and (MoNbTaVW)C subject to elongation parallel to low-index crystallographic directions [001], [110], and [111] are presented in **Fig. 3**. Major emphasis is put on the results obtained for strained B1 HEC(001) supercells, **Fig. 3a**, because (001) planes are expected to be the easiest cleavage planes in inherently brittle B1 carbonitrides [62]. Note that the stress-strain responses described here by AIMD simulations represent localized deformation mechanisms and hence are not necessarily predictive of bulk material responses. Nevertheless, AIMD simulations of defect-free models subject to deformation enable realistic predictions of potential plasticity routes occurring within crystal grains of real alloys.

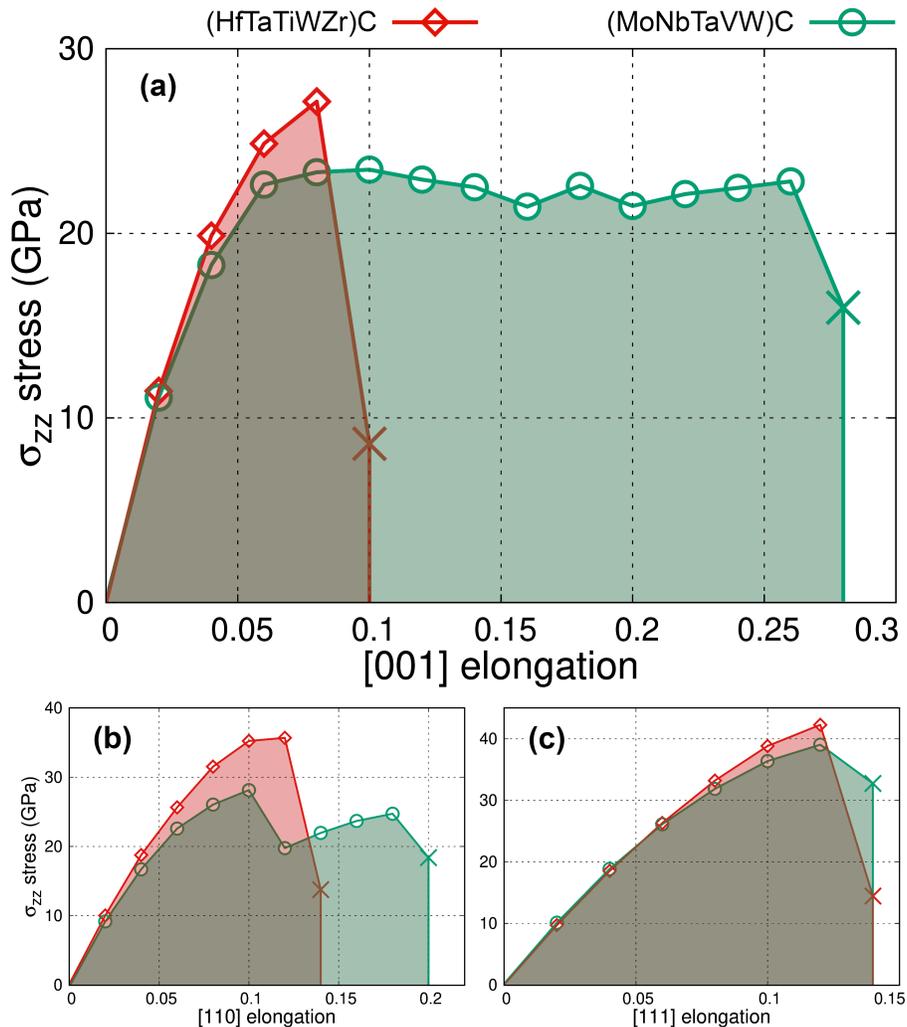

**Figure 3.** Stress/strain curves calculated for (HfTaTiWZr)C (red diamonds) and (MoNbTaVW)C (green circles) subject to tensile strain along **(a)** the [001], **(b)** the [110], and **(c)** the [111] crystallographic directions at room temperature. The "×" indicate fracture points.

AIMD simulations indicate that, for [001] elongations ≲6%, both B1 HEC(001) crystals exhibit a similar (quasi)-elastic response to strain. The $C_{11}$ elastic stiffnesses, calculated from the slope of $\sigma_{zz}$ vs. δ for δ → 0, are nearly identical (≈615-620 GPa) [50]. The two carbide systems reach their yield point for elongations δ ~ 8%. The calculated ideal tensile strengths are $\sigma_T^{(HfTaTiWZr)C\ (001)}$ = 27 GPa and $\sigma_T^{(MoNbTaVW)C\ (001)}$ = 23 GPa, with corresponding moduli of resilience $U_R^{(HfTaTiWZr)C\ (001)}$ = 1.4 GPa and $U_R^{(MoNbTaVW)C\ (001)}$ = 0.8 GPa.

The most remarkable difference in the mechanical behavior of the two multicomponent carbides, subject to [001] strain, is beyond their yield points. **Fig. 4a,b** shows that both (MoNbTaVW)C (001) and (HfTaTiWZr)C (001) retain an ideal octahedral coordination up to a strain of 6%. An elongation of 8%, which corresponds to the maximum extension withstood by (HfTaTiWZr)C (001), induces puckering of some atomic layers (note wavy lattice planes in **Fig. 4b**). Puckering of lattice planes is also observed in (MoNbTaVW)C strained by 8 or 10% (**Fig. 4a**). In contrast, an increase in strain to 10% is sufficient for opening a crack in (HfTaTiWZr)C (001), which rapidly causes cleavage of the material on a (001) plane (see simulation snapshots in **Fig. 4b** and corresponding sudden drop in the $\sigma_{zz}$ stress of (HfTaTiWZr)C in **Fig. 3a**; **Video1** in the **SM**). By contrast (MoNbTaVW)C (001), exhibits considerable resistance to fracture. Elongated by 12% or more, this high entropy carbide undergoes lattice transformations, with local modifications in the bonding network and atomic coordination, which allows for dissipation of a portion of the accumulated stress. The transformation toughening mechanism retards (MoNbTaVW)C (001) fracture, which occurs at a total elongation of 28%. The process consists in a relatively slow fraying of chemical bonds (see stress/strain curve in **Fig. 3a** and simulation snapshots on the right of **Fig. 4a**; **Video2** in the **SM**). Owing to strain-induced transformations, the total ideal toughness (derived from these small cells, not necessarily indicative of bulk toughness) calculated for (MoNbTaVW)C (001) ($U_T^{(MoNbTaVW)C\ (001)}$ = 5.7 GPa) is more than three times that obtained for (HfTaTiWZr)C (001) ($U_T^{(HfTaTiWZr)C\ (001)}$ = 1.8 GPa).

To gain a more detailed understanding of ideal strength, local plasticity, and their potential influence on resistance to fracture in pristine (defect-free) (MoNbTaVW)C and (HfTaTiWZr)C, AIMD simulations are also performed for HEC(110) and HEC(111) supercells strained along the [110] and [111] axes, respectively. The results of corresponding stress/strain curves up to fracture are illustrated in **Fig. 3b** and **3c**. As for the case of HEC(001) crystals, (HfTaTiWZr)C elongated parallel to [110] and [111] displays ≈10–20% greater tensile strengths $\sigma^{[111]}$ and $\sigma^{[110]}$ than (MoNbTaVW)C (see **Fig. 3b,c** and **Table 1**). The higher strength of (HfTaTiWZr)C may explain its higher hardness values H in comparison to (MoNbTaVW)C ($H^{(HfTaTiWZr)C}$ =33±2 GPa vs. $H^{(MoNbTaVW)C}$ = 27±3 GPa) [6]. Indeed, the hardness of a solid is typically correlated to its mechanical strength [64,65].

The stress/strain results illustrated in **Fig. 3b** demonstrate that the response of the two HEC systems to [110] strain is similar to that observed upon tensile deformation along [001]. While (HfTaTiWZr)C (110) rapidly cleaves beyond its yield point (note stress drop at 14% strain of (HfTaTiWZr)C (110) in **Fig. 3b**, see also **Video3** in the **SM**), (MoNbTaVW)C (110) activates a local lattice transformation, which inhibits brittle fracture and enhances the ideal crystal's potential plasticity routes. AIMD simulations indicate that (MoNbTaVW)C (110) can withstand a relative elongation of 20%, that produces a total toughness $U_T^{(MoNbTaVW)C\ (110)} = 4.1$ GPa. The toughness calculated for brittle (HfTaTiWZr)C (110) is ≈25% lower: $U_T^{(HfTaTiWZr)C\ (110)} = 3.3$ GPa. The strain-induced structural transformation of (MoNbTaVW)C (110) is presented in **Fig. 5**. Upon [110] strain, (MoNbTaVW)C maintains an octahedral B1-like atomic coordination up to a strain of 10% (**Fig. 5a**), which corresponds to the material yield point (**Fig. 3b**). Further [110] elongation to 12% induces a local lattice transformation with formation of a graphitic-like crystal structure (**Fig. 5b**, **Video4** in the **SM**). Electronic densities of states (not shown) directly extracted by AIMD simulations at 300K show that both unstrained B1-structure (MoNbTaVW)C and 18%-elongated B1/graphite-like (MoNbTaVW)C are metallic systems (finite DOS at the Fermi level). The B1/graphite phase mixture sustains elongation up to 20%, when it fractures via relatively slow (≈2 ps) bond fraying at the interface between the B1 and graphitic-like domains (**Fig. 5d**).

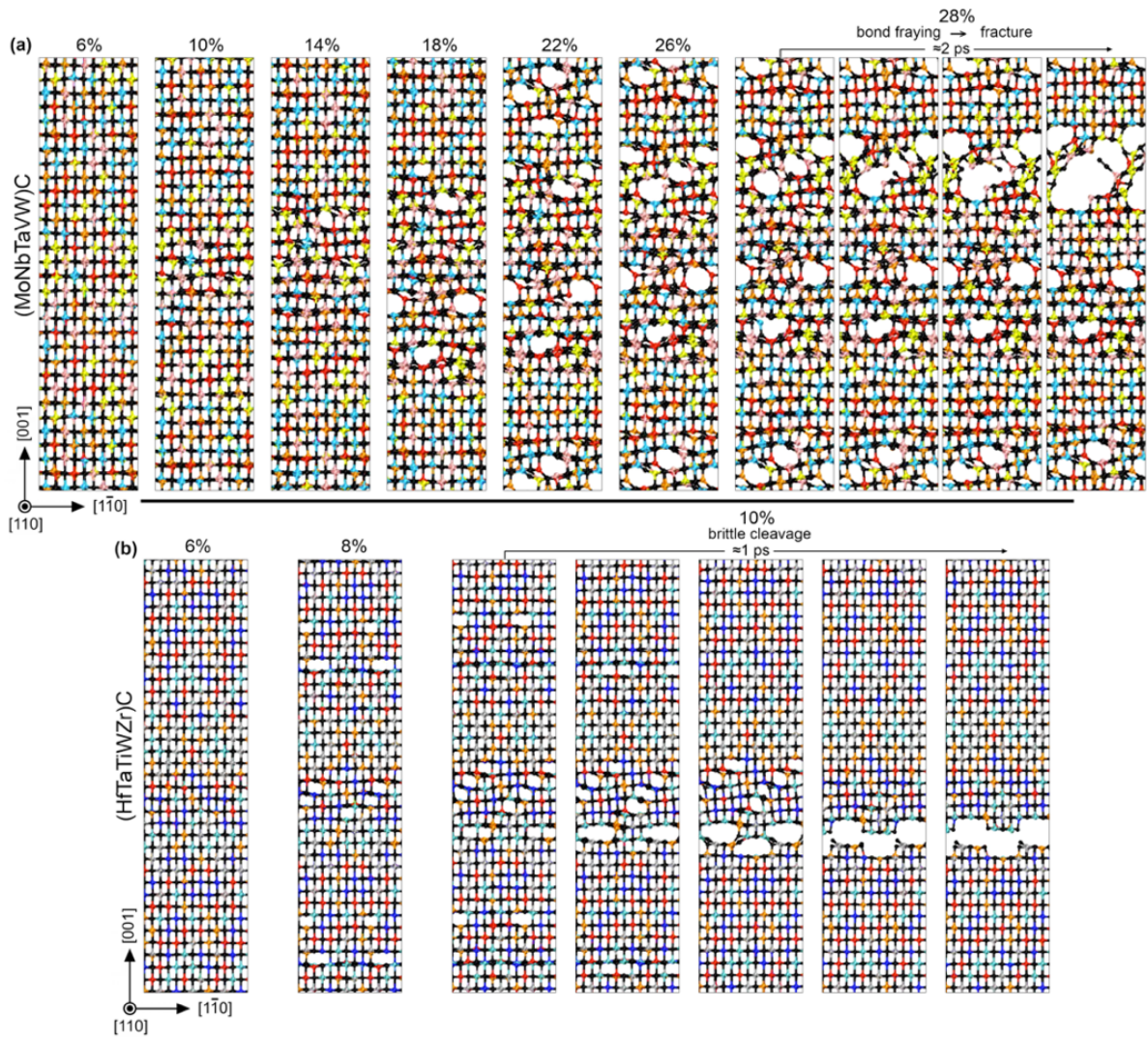

**Figure 4.** Orthographic view of AIMD snapshots during [001] elongation of **(a)** (MoNbTaVW)C (**Video2** in the **SM**) and **(b)** (HfTaTiWZr)C (**Video1** in the **SM**). Beyond its yield point (≈8%), (MoNbTaVW)C undergoes severe lattice transformations, which effectively retard fracture up to an elongation of 28%. Conversely, (HfTaTiWZr)C fractures in a brittle manner by sudden bond snapping at a strain of 10%. Color legend for atomic spheres in (MoNbTaVW)C: black = C, pink = Nb, yellow = Mo, red = Ta, light blue = V, orange = W and (HfTaTiWZr)C: black = C, silver = Ti, cyan = Zr, red = Ta, blue = Hf, orange = W. Dynamic bonds have cutoff lengths of 2.6 Å.

AIMD simulations show that the (HfTaTiWZr)C and (MoNbTaVW)C defect-free models possess a similar mechanical response to [111] strain. In this case, both materials fail in a brittle manner at an elongation of 14% (see **Videos 5,6** in the **SM**). However, fracture occurs preferentially on (001) crystal facets following a zig-zag pattern (analogous to crack opening in B1 TiN(111), see figure 10 in Ref. [56]). The total ideal toughness values calculated for (HfTaTiWZr)C and (MoNbTaVW)C upon elongation parallel to [111] are 3.5GPa. In this case, the total toughness is essentially equivalent to the moduli of resilience (fracture occurs directly after mechanical yielding). AIMD results of ideal strengths $\sigma_T$, moduli of resilience $U_R$, total toughness $U_T$ and elongation at fracture $\delta_f$ are summarized in **Table 1**.

| | (HfTaTiWZr)C | | | | | | (MoNbTaVW)C | | | | | |
|---|---|---|---|---|---|---|---|---|---|---|---|---|
| Elongation | $\delta_u$ (%) | $\sigma_T$ (GPa) | $U_R$ (GPa) | $U_T$ (GPa) | $\delta_f$ (%) | mechanism | $\delta_u$ (%) | $\sigma_T$ (GPa) | $U_R$ (GPa) | $U_T$ (GPa) | $\delta_f$ (%) | mechanism |
| <001> | 8 | 27 | 1.4 | 1.8 | 10 | brittle cleav. | 10 | 23 | 0.8 | 5.7 | 28 | transf.→ductile fract. |
| <110> | 12 | 35 | 2.1 | 3.3 | 14 | brittle cleav. | 10 | 28 | 1.8 | 4.1 | 20 | transf.→ductile fract. |
| <111> | 12 | 42 | 2.9 | 3.5 | 14 | brittle cleav. | 12 | 39 | 2.8 | 3.5 | 14 | brittle cleavage |

**Table 1.** AIMD-calculated properties for defect-free B1 (HfTaTiWZr)C and (MoNbTaVW)C subject to [001], [110], and [111] uniaxial strain. $\delta_u$ is the elongation in correspondence of the ultimate tensile strength $\sigma_T$ value. $U_R$ is the modulus of resilience (elastic-strain energy density accumulated up to yielding), $U_T$ the total ideal toughness (strain energy density accumulated before fracture), and $\delta_f$ is the elongation at fracture.

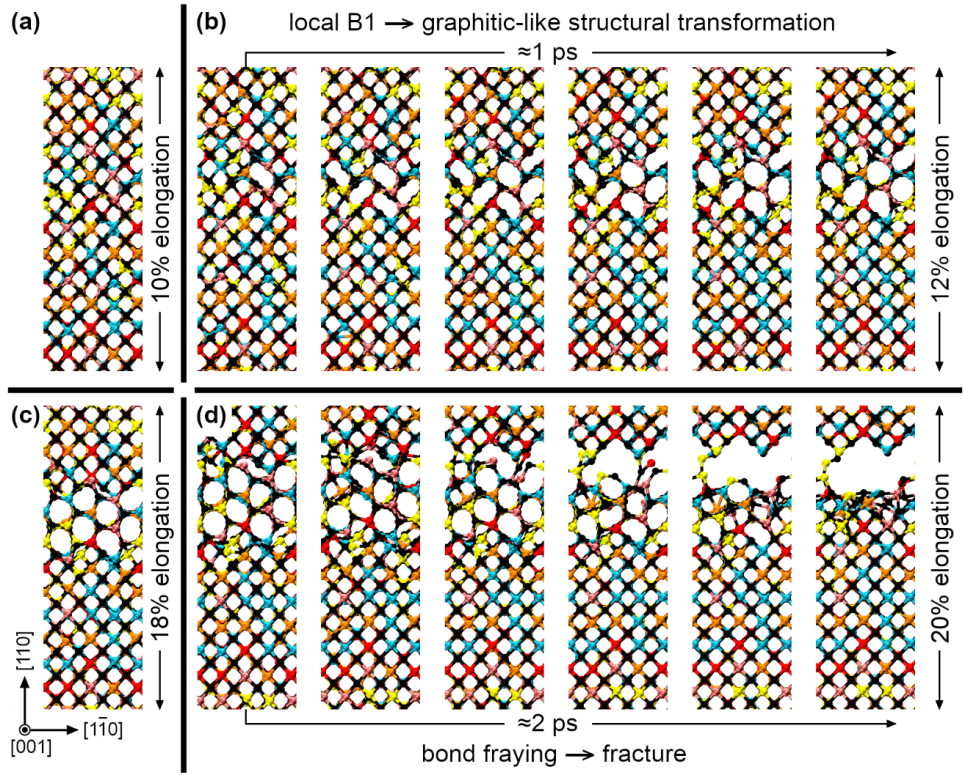

**Figure 5.** Orthographic view of AIMD snapshots during [110] elongation of (MoNbTaVW)C (**Video4** in the **SM**). **(a)** The material maintains octahedral atomic coordination up to a strain of 10%. **(b)** An elongation of 12% activates lattice transformations characterized by a local transition from B1 to a graphitic-like structure. **(c)** The B1/graphite superlattice structure withstands strains up to 18%. **(d)** Fracture occurs via a progressive bond fraying at 20% elongation. Color legend for atomic spheres in (MoNbTaVW)C: black = C, pink = Nb, yellow = Mo, red = Ta, light blue = V, orange = W. Dynamic bonds have cutoff lengths of 2.6 Å.

Simulations are also employed to assess the intrinsic resistance of (HfTaTiWZr)C and (MoNbTaVW)C crystals to slip along $\{001\}\langle 1\bar{1}0\rangle$, $\{110\}\langle 1\bar{1}0\rangle$, $\{111\}\langle 1\bar{1}0\rangle$, and $\{111\}\langle 11\bar{2}\rangle$. The $\{110\}\langle 1\bar{1}0\rangle$ and $\{111\}\langle 1\bar{1}0\rangle$ slip systems are typically operative in B1 structured carbides at low, moderate, or high temperatures [35,66]. In contrast, $\{001\}\langle 1\bar{1}0\rangle$ slip has been experimentally reported in very few cases [67,68]. The modeling of $\{111\}\langle 11\bar{2}\rangle$ shearing allows assessing the carbide ability to form {111} stacking faults or twin boundaries upon loading [69]. The shear-stress $\sigma_{xz}$ vs. trigonal strain on $\{001\}\langle 1\bar{1}0\rangle$, $\{110\}\langle 1\bar{1}0\rangle$, $\{111\}\langle 1\bar{1}0\rangle$, and $\{111\}\langle 11\bar{2}\rangle$ slip systems calculated for (HfTaTiWZr)C and (MoNbTaVW)C are illustrated in **Fig. 6** (see shear deformation of (MoNbTaVW)C at 300 K in **Videos 7-10** in the **SM**). Ideal shear strengths $\gamma_S$ and deformation at slip $\delta_S$ values are summarized in **Table 2**.

For all shear deformations considered here, slip occurs according to Schmid's law, that is, the observed glide plane is the one on which the resolved shear stress is highest (see **Videos 7-10** in the **SM**). In this regard, it is worth underlining that previous *ab initio* calculations showed that B1 pseudobinary (Ti,W)N refractory ceramics may exhibit non-Schmid slip behavior (figures 9 and 10 in [70]). For the investigated HEC systems, anomalous non-Schmid slip is observed during $\{111\}\langle 1\bar{1}0\rangle$ shearing at 900 and 1200K (**Videos 11,12** in the **SM**). The calculated ideal strengths $\gamma_S$ of (MoNbTaVW)C are lower than those obtained for (HfTaTiWZr)C (**Fig. 6** and **Table 2**). (MoNbTaVW)C also displays a lower shear modulus $C_{44}$ (145GPa) [50] in comparison with (HfTaTiWZr)C (165 GPa) [50], as reflected by a lower initial slope of $\sigma_{xz}$ vs. $\{001\}\langle 1\bar{1}0\rangle$-strain (**Fig. 6a**). In addition, the shear moduli $G_{110}$ and $G_{111}$ (resolved on $\{110\}\langle 1\bar{1}0\rangle$ and $\{111\}\langle 1\bar{1}0\rangle$ slip systems) evaluated for (HfTaTiWZr)C (248 and 220 GPa, respectively) [50] are greater than for (MoNbTaVW)C (229 and 201 GPa, respectively) [71].

A lower shear resistance in (MoNbTaVW)C is expected due to its higher occupation of *d-d* electronic states, which enables formation of metallic bonds upon shearing [39,40,72]. In addition, the smaller shear stresses $\gamma_S$ required to induce slip in (MoNbTaVW)C (**Table 2**), are expected to reflect trends in Peierls stresses necessary to move dislocations in real carbide systems. Overall, the lower intrinsic *lattice friction*, as reflected by lower $\gamma_S$ values calculated for defect-free (MoNbTaVW)C, is likely to further contribute to prevent brittle fracture, as clarified in **Section 4**. It should be noted that realistic *ab initio* calculations of dislocation core structures, dislocation glide, and corresponding Peierls stresses, in binary ceramics are feasible. For example, *ab initio* results for B1 TiN and rutile $TiO_2$ are discussed in Yadav *et al.* [73] and Maras *et al.* [74]. Nevertheless, it has also been demonstrated by experiments and calculations that the dislocation cores in simple B1 MgO binary ceramic systems present different non-intuitive polymorph structures [75]. It is plausible to assume that determination of dislocation core structures in high-entropy ceramic systems is considerably more challenging than in binaries due to different chemical species involved.

| Shearing | (HfTaTiWZr)C | | (MoNbTaVW)C | |
|---|---|---|---|---|
| | $\gamma_S$ (GPa) | $\delta_S$ (%) | $\gamma_S$ (GPa) | $\delta_S$ (%) |
| {001}<1-10> | 21.9 | 16 | 18.6 | 18 |
| {110}<1-10> | 27.9 | 14 | 21.8 | 12 |
| {111}<1-10> | 28.4 | 16 | 21.1 | 14 |
| {111}<11-2> | 22.0 | 14 | 20.0 | 14 |

**Table 2.** AIMD results at 300K for ideal shear strengths $\gamma_S$ and shear-strain $\delta_S$ required to induce slip in dislocation-free B1 (HfTaTiWZr)C and B1 (MoNbTaVW)C crystals on four different slip systems.

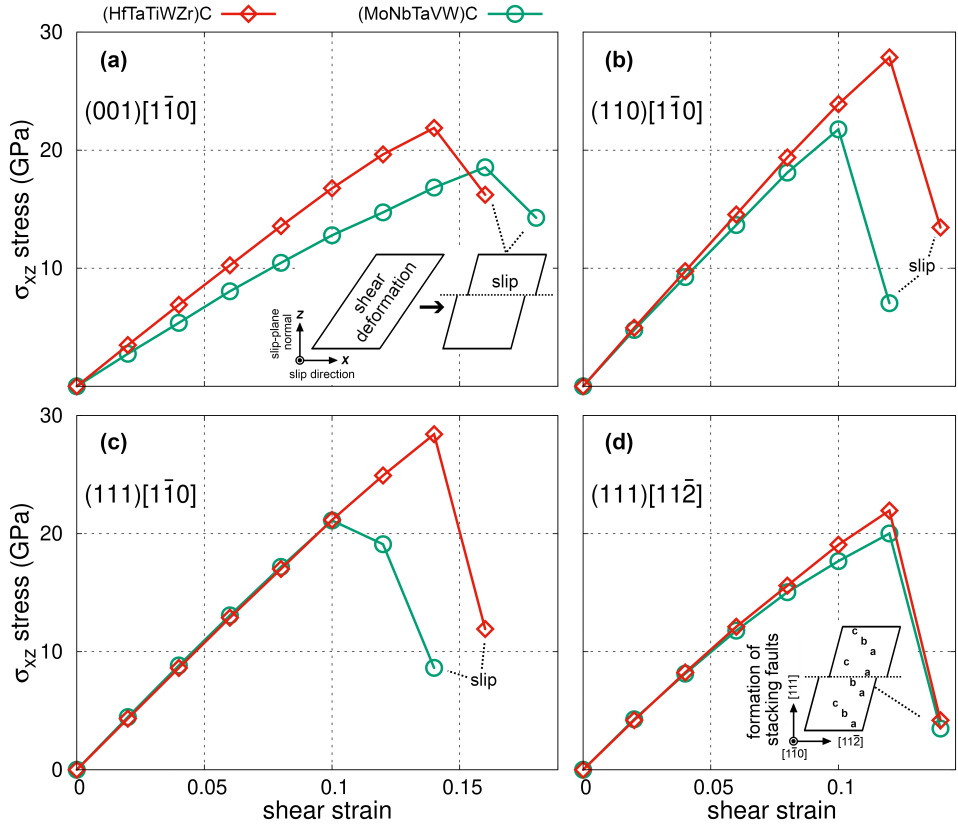

**Figure 6.** AIMD stress vs. strain curves calculated for **(a)** {001}⟨1̄10⟩, **(b)** {110}⟨1̄10⟩, **(c)** {111}⟨1̄10⟩, and **(d)** {111}⟨112̄⟩ shearing in (HfTaTiWZr)C and (MoNbTaVW)C. The insets in **(a)** and **(d)** are schematic illustrations of lattice slip (after which octahedral coordination is restored) and formation of stacking faults, which locally modify the stacking sequence of a B1 lattice from -a-b-c- to a-b-a-c. The shear strength ($\gamma_S$) corresponds to the maximum stress $\sigma_{xz}$ withstood before occurrence of lattice slip **(a-c)** or formation of stacking faults **(d)**. **Videos 7-10** in the **SM**, are examples of lattice slip and formation of {111} stacking faults at 300K in (MoNbTaVW)C.

## 4. Discussion

Nanoindentation results of this work suggest that (MoNbTaVW)C is considerably more resistant to fracture than (HfTaTiWZr)C (**Fig. 2**). Since the stress field generated under a Berkovich indenter comprises tensile stresses at the indenter corners [76], where radial (Palmqvist) fracture occurs, uniaxial tensile deformation modeled via AIMD for pristine crystal models provides useful information to clarify the differences in observed response of (HfTaTiWZr)C and (MoNbTaVW)C under indentation load. The transformations induced in (MoNbTaVW)C during tensile elongation may hinder the formation and/or propagation of cracks via modification of the bonding network at the crack front. Moreover, AIMD results show that lattice slip in (MoNbTaVW)C is activated at lower shear stresses than in (HfTaTiWZr)C (**Fig. 6**). The relatively lower resistance to lattice slip (**Fig. 6a-c**) and more facile formation of stacking faults (**Fig. 6d**) demonstrated for ideal (MoNbTaVW)C crystals may provide additional pathways for stress dissipation at a crack tip (i.e., enhanced plasticity) [77].

As discussed below, the electron concentration of the alloy and preferred lattice structure of binary constituents are key parameters to design high-entropy ceramics with enhanced hardness and toughness. In this work, (HfTaTiWZr)C and (MoNbTaVW)C are selected as representative high-entropy carbide systems due to their specific VEC values. The electron concentration of (HfTaTiWZr)C is expected to provide very high hardness. In contrast, the VEC of (MoNbTaVW)C is optimal for achieving a material with both high hardness and enhanced plasticity routes that potentially can lead to improved toughness.

In B1 (HfTaTiWZr)C, the valence electrons (8.6 e$^-$/f.u.) fill nearest-neighbor carbon (p) – metal (d) electronic states, which provide strong bonding, while leaving the shear-sensitive metallic states unoccupied [72]. Such an electronic mechanism contributes to the explanation of why (HfTaTiWZr)C displays the highest measured hardness (33 GPa) in relation to several other single-phase B1 high-entropy carbides [6]. It is also worth noting that the higher hardness of (HfTaTiWZr)C in relation to (MoNbTaVW)C (27 GPa) is reflected in higher ideal tensile and shear strengths, as well as higher elastic and shear moduli $C_{44}$, $G_{110}$, and $G_{111}$ [51]. On the other hand, high hardness alone does not guarantee good resistance to fracture. AIMD simulations show that (HfTaTiWZr)C systematically cleaves at its yield point when subjected to tensile strain (**Fig. 3** and **Fig. 4b**). In fact, ceramics are typically brittle materials, prone to crack when subjected to excessive load [78]. Nonetheless, predictions of previous *ab initio* investigations [39], later validated by experiments [46], demonstrated that it is possible to design *inherently* hard and tough single-crystal B1 pseudobinary nitride ceramics via electronic-structure manipulation. The toughness is expected to be maximized for VEC ≈10.5 e$^-$/f.u. due to full occupation of *d-d* bonding states [45], which, in turn, favor activation of plastic deformation at the material yield point [60].

Transition-metal carbides have one electron less than nitrides for a given composition of refractory elements. Thus, the VEC (9.4 e$^-$/f.u.) of (MoNbTaVW)C samples is optimal to probe the effects induced by an increased occupation of *d-d* metallic states on the mechanical behavior of the material. In addition to improving the metallic character of the ceramic, the VEC and specific cation composition of (MoNbTaVW)C may enhance hardness and toughness as well, due to a favored formation of stacking faults. Indeed, Hugosson *et al.* [79] showed that a VEC near 9.5 e$^-$/f.u. sets the B1 and hexagonal structures of pseudobinary carbides to equal energies, thus promoting nucleation of stacking faults, which increases the material hardness by restricting dislocation motion across the fault. As a proof of concept, Joelsson *et al.* characterized pseudobinary B1 $Zr_{1-x}Nb_xN$ ceramics as a function of the metal composition [80]. The authors demonstrated that the material reaches its maximum hardness for x≈0.5 (VEC≈9.5 e$^-$/f.u.) and attributed the effect to the observation of an increased density of stacking faults. Our previous nanoindentation measurements revealed that (MoNbTaVW)C possesses high hardness (≈27 GPa) [6]. This may be due to a high concentration of intrinsic stacking faults. Note that high hardness is an important prerequisite for achieving increasingly tougher ceramics.

Other recent *ab initio* investigations showed that facile nucleation of {111} stacking faults in B1 structured ceramics may promote ductility (beside hardness [79]) by assisting slip *along* the fault [71]. The mixing of binary transition-metal nitride and/or carbide systems – one with B1 and the other with hexagonal structures – yields metastable B1 compounds with an ability to glide on {111} crystallographic planes at relatively low temperatures. The latter deformation pathway is important for achieving good ductility, according to the criteria proposed by von Mises [81]. The mechanism is owed to an energetically-favorable formation of {111} stacking faults in domains rich in group-VIB transition-metal nitrides or carbides, which are thermodynamically-inclined to crystallize in the WC-structure with A-B-A-B stacking sequence [71] or other hexagonal structures [79]. Moreover, 0-K *ab initio* results by Yu *et al*. showed that group-VB carbides (VEC=9 e⁻/f.u.) exhibit an energetic preference to nucleate stacking faults during {111}⟨11$\bar{2}$⟩ slip [82]. In contrast, {111} stacking faults are predicted to be unstable in group-IVB carbides (VEC = 8 e⁻/f.u.) at 0K [82]. To summarize, carbides with VEC = 8 e⁻/f.u. are less energetically inclined to activate lattice slip in general, in comparison with carbides with VEC ≥ 9 e⁻/f.u. (see figure 3 in Ref. [82]). The fact that the (MoNbTaVW)C system comprises 40% metal concentration of Mo and W and the remaining 60% of group-VB transition-metals explains the lower shear stresses $\gamma_S$ required to activate lattice slip (**Fig. 6a,b,c**) or form {111} stacking faults during {111}⟨11$\bar{2}$⟩ shearing (**Fig. 6d**) in relation to (HfTaTiWZr)C.

The energetic inclination of B1 carbides with VEC ≥ 9 e⁻/f.u. to slip on {111} planes at room temperature [66] and/or nucleate {111} stacking faults [70,82] originates from *d-d* electron hybridization activated with shear deformation. The shear-induced enhancement of the metallic character of the material – evidenced by *d*-electron accumulation between metal atoms: 2$^{nd}$- neighbor [39,47,71] or 4$^{th}$-neighbors (across anion vacancy sites) [61] – assists lattice slip via electron reorganization along the glide plane. However, DFT charge-density results of this work show that a high VEC can also enable modifications in atomic coordination during tensile loading. The effects of such an electronic mechanism are illustrated in **Fig. 7**. The figure offers a comparison between the electron-transfer maps calculated for (HfTaTiWZr)C (001) and (MoNbTaVW)C (001) elongated by 10 and 18%, respectively. The onset of brittle fracture in tensile-strained (HfTaTiWZr)C (001) is manifested by the formation of an electron-depleted region (yellow color) between atomic layers (**Fig. 7a**). Cleavage of the crystal is slightly delayed by a few metal/carbon chemical bonds able to bend diagonally across the electron-depleted region (see dashed lines in **Fig. 7a**). In contrast, the high density of metallic electronic states of (MoNbTaVW)C allows the material to modify bond angles and atomic coordination, which dissipates accumulated stresses at high load conditions. Elongation beyond the yield point of (MoNbTaVW)C (001) triggers electron transfer, which in turn, enables local lattice transformations. The relatively packed B1 structure becomes more open: many atoms change coordination from six to five (as marked, e.g., for two Nb and Mo atoms in **Fig. 7b**) as the bond angles within the (100) plane seen in the figure increase from 90 to 120º. Remarkably, during [110] elongation, a high VEC allows a significant portion of the (MoNbTaVW)C lattice to transform from six-fold coordinated B1 to a fivefold-coordinated graphite-like structure, as illustrated in **Fig. 5**.

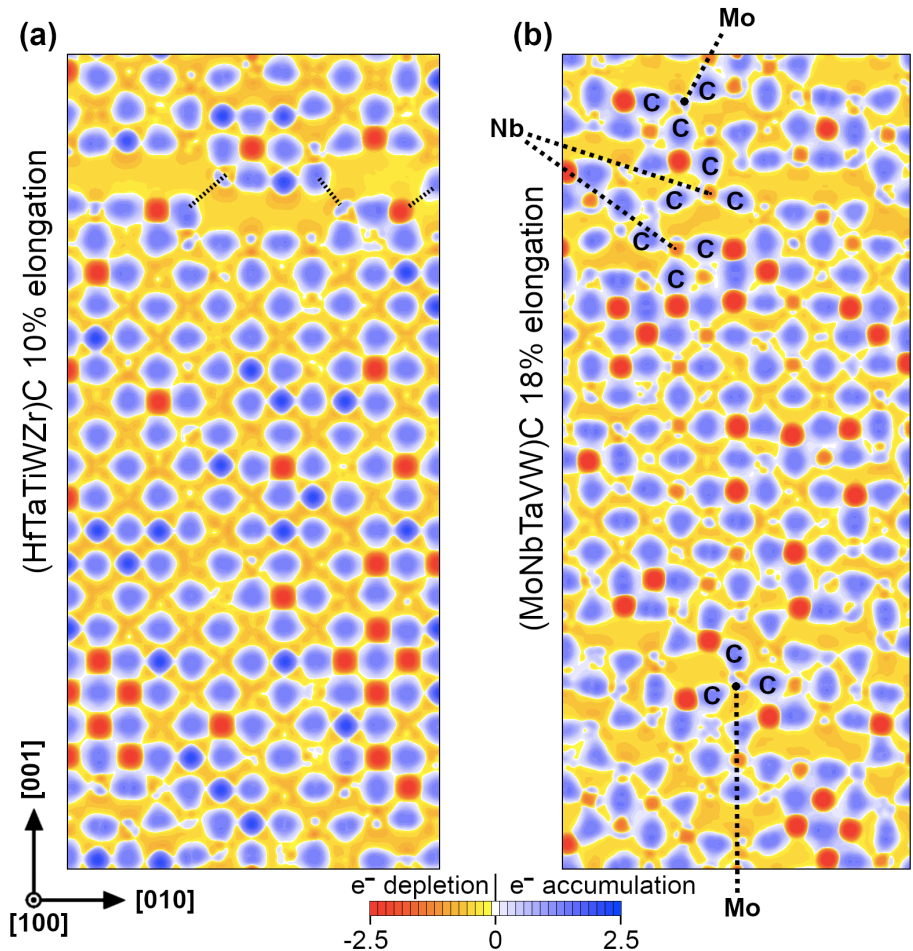

**Figure 7.** Electron transfer maps sliced on a (100) lattice plane of (HfTaTiWZr)C (001) **(a)** and (MoNbTaVW)C (001) **(b)** supercells subject to 10 and 18% uniaxial elongation, respectively. Note that the crystal lattice of (HfTaTiWZr)C in panel **(a)** is still unbroken: bonds across the electron-depleted region are indicated by a dotted line. The electron density in **(a)** is calculated for the (HfTaTiWZr)C supercell configuration shown on the **fourth panel of Fig. 4b**. The color scale is expressed in $e^-/Å^3$.

Additional AIMD simulations of mechanical testing indicate that the enhanced plasticity and potentially increased resistance to fracture of (MoNbTaVW)C, here computationally demonstrated at 300K, may be preserved up to temperatures between 900 and 1200K. A comparison of the results of AIMD simulations for (HfTaTiWZr)C and (MoNbTaVW)C at 600, 900, and 1200K, performed for the same tensile and shear strain paths shown in **Tables 1 and 2**, are presented in the **SM**. The most important findings are summarized below. Remarkably, both materials exhibit very slow reductions in tensile and shear strengths with temperature (**Figs. 1S,2S**). Parallel AIMD investigations conducted on binary and pseudobinary refractory ceramic systems – for which both the tensile and shear strengths decrease monotonically with increasing temperature – suggest that the mechanical strength of high-entropy carbide alloys is relatively unaffected by temperature. Concerning plasticity and resistance to fracture, AIMD results indicate that (HfTaTiWZr)C remains relatively hard, but brittle at all investigated temperatures: elongated up to its yield point, the material cleaves via rapid bond snapping (**Tables 1S,3S**). Exception is made for [001] elongation at 600K, for which (HfTaTiWZr)C fractures via slow fraying of chemical bonds at 14% strain.

Conversely, (MoNbTaVW)C maintains outstanding toughness and resistance to brittle fracture during [001] and [110] elongation up to 900K (**Tables 2S,3S**; **Videos 13,14**). This is manifested by lattice transformations, which lead to significant plastic deformation beyond the material yield point. At 1200K, the resistance to fracture of (MoNbTaVW)C appears to have degraded: the crystal cleaves upon [110] elongation without forming graphite-like networks. Nevertheless, the material still presents a modest resistance to brittle failure during [001] elongation also at 1200K: the yield strength is met at 8% strain, whereas fracture occurs via slow bond fraying at 14% strain (**Video 15**).

## 5. Conclusions

The mechanical response of high entropy carbides was investigated experimentally and through AIMD simulations. The two compositions (HfNbTaTiZr)C and (MoNbTaVW)C with valence electron concentrations of 8.6 $e^-$/f.u. and 9.4 $e^-$/f.u., respectively, were chosen to demonstrate noticeable differences in plasticity and resistance to fracture. Experimental observations of fracture frequency in indentation arrays at various loads and depths demonstrate that the fracture resistance of the more electron rich (MoNbTaVW)C exceeds that of (HfNbTaTiZr)C. All indentations into the (HfNbTaTiZr)C surface exhibit fracture with as little as 250 mN force (or 600 nm depth), while indents into the (MoNbTaVW)C surface consistently do *not* exhibit fracture up to 500 mN force (or 1,000 nm depth). AIMD simulations of tensile deformation along the [001], [110], and [111] crystal axes of defect-free supercell models are utilized to corroborate the observed differences in ductility and toughness. From these results, it can be gleaned that the increased resistance to fracture in (MoNbTaVW)C may originate from modifications in the bonding network triggered at the material yield point. The transformation toughening mechanisms revealed for tensile-strained pristine (MoNbTaVW)C vary depending on the loading direction, but necessarily entail a reduction in the atomic coordination. For example, subject to [110] elongation, the sixfold-coordinated B1 lattice locally transforms to a fivefold-coordinated graphite-like structure. Furthermore, the AIMD simulations show that the *d-d* electron hybridization activated with shear deformation in the carbide with VEC $\geq$ 9 $e^-$/f.u. allows for an energetic preference to form {111} stacking faults, which may allow for further stress dissipation at crack tips through the creation of new plasticity routes, thus positively contributing to the material's resistance to fracture. Complementary AIMD modeling of mechanical properties as a function of temperature indicate that both (HfNbTaTiZr)C and (MoNbTaVW)C exhibit very slow reductions in tensile and shear strengths with temperature and that (MoNbTaVW)C may preserve good fracture resistance up to 900K. This work offers a simple method, via VEC tuning in high entropy carbides, for the design of refractory ceramic materials with inherently improved plasticity and fracture resistance at room and elevated temperatures.

**Acknowledgements**

All simulations were carried out using the resources provided by the Swedish National Infrastructure for Computing (SNIC) – partially funded by the Swedish Research Council through Grant Agreement Nº VR-2015-04630 – on the Clusters located at the National Supercomputer Centre (NSC) in Linköping, the Center for High Performance Computing (PDC) in Stockholm, and at the High Performance Computing Center North (HPC2N) in Umeå, Sweden. This work was partially supported through access and utilization of the UC San Diego, Dept. of NanoEngineering's Materials Research Center. D.G.S. gratefully acknowledges financial support from the Competence Center Functional Nanoscale Materials (FunMat-II) (Vinnova Grant No. 2016–05156) and the Olle Engkvist Foundation. The authors K.S.V., T.H., and W.M. acknowledge support through the Office of Naval Research ONR-MURI (grant No. N00014-15-1-2863). K.K. acknowledges support from the Department of Defense (DoD) through the National Defense Science and Engineering Graduate (NDSEG) Fellowship Program. K. K. would also like to acknowledge the financial support of the ARCS Foundation, San Diego Chapter.

**Credit authorship contribution statement**

**Davide G. Sangiovanni:** Conceptualization, Methodology, Software, Validation, Formal analysis, Writing – Original Draft, Writing – Review & Editing, Visualization, Funding acquisition. **William Mellor:** Methodology, Validation, Investigation, Writing – Original Draft, Writing – Review & Editing, Visualization. **Tyler Harrington:** Conceptualization, Methodology, Writing – Review & Editing. **Kevin Kaufmann:** Methodology, Validation, Formal analysis, Investigation, Writing – Review & Editing, Visualization. **Kenneth Vecchio:** Conceptualization, Methodology, Validation, Writing – Original Draft, Writing – Review & Editing, Supervision, Project administration, Funding acquisition.